\documentclass[aip,jcp,preprint,amsmath,amssymb, groupedaddress]{revtex4-1}
\usepackage[T1]{fontenc}
\usepackage[utf8]{inputenc}
\usepackage{amsfonts}
\usepackage{multirow,dcolumn}
\usepackage{booktabs}
\usepackage[version=3]{mhchem}
\usepackage{setspace} 
\usepackage{lmodern}

\usepackage{natbib}
\usepackage[pdftex]{graphicx}
\usepackage[amssymb]{SIunits} 

\usepackage{paralist} 
\usepackage[normalem]{ulem}

\newcommand{\dr}{(2)}
\newcommand{\tr}{(3)}
\newcommand{\dg}[1]{#1^\dagger}
\newcommand{\dgg}{^\dagger}
\newcommand{\esd}{e^{S^\dagger}}
\newcommand{\esdm}{e^{-S^\dagger}}

\newcommand{\etm}{e^{-T}}

\newcommand{\et}{e^{T}}
\newcommand{\comute}[2]{ [ #1,#2]}
\newcommand{\cm}[2]{ [ #1,#2]}
\newcommand{\rar}{\rightarrow}

\newcommand{\cmm}[3]{ [[ #1,#2], #3]}
\newcommand{\cmmb}[3]{ [#1, [ #2,#3]]}
\newcommand{\sump}{\sideset{}{'}\sum}

\newcommand{\Ome}[2]{\Omega_{#1}^{#2}(\omega)}
\newcommand{\Omx}{\Omega^X(\omega)}
\newcommand{\odp}{\langle\langle X;Y\rangle\rangle _\omega}
\newcommand{\odpm}{\langle\langle X;Y\rangle\rangle _{-\omega}}

\newcommand{\fr}[1]{Eq.~(\ref{#1})}
\newcommand{\frm}[1]{Eqs.~(\ref{#1})}
\newcommand{\Frff}[2]{Figs.~\ref{#1}~to~\ref{#2}}
\newcommand{\Frt}[1]{Table~\ref{#1}}
\newcommand{\frs}[1]{section~\ref{#1}}

\newcommand{\mt}[1]{\mbox{\tiny{#1}}}
\newcommand{\relerror}{$|\delta q|/|q_\mathrm{exp}| \times 100 \% $}
\newcommand{\errname}{$\Delta_\mathrm{rel}$}
\newcommand{\bigoh}[1]{the #1 order}

\newcommand{\braket}[2]{\left\langle #1\middle|#2\right\rangle} 
\newcommand{\brakett}[3]{\left\langle #1\middle #2|#3\right\rangle} 
\newcommand{\brz}[1]{\left\langle #1\right\rangle} 

\newcommand{\equ}[1]{\begin{equation} #1 \end{equation}}
\newcommand{\equl}[2]{\begin{equation}\label{#2} #1 \end{equation}}
\newcommand{\equm}[1]{\begin{multline} #1 \end{multline}}
\newcommand{\equml}[2]{\begin{multline}\label{#2} #1 \end{multline}}

\newcommand{\equal}[2]{\begin{align}\label{#2} #1 \end{align}}
\newcommand{\equs}[1]{\begin{equation}\begin{split} #1 \end{split}\end{equation}}
\newcommand{\equsl}[2]{\begin{equation}\begin{split}\label{#2} #1 \end{split}\end{equation}}
\begin{document}

\title{Transition properties from the Hermitian formulation of the
 coupled cluster polarization propagator}

\author{Aleksandra M. Tucholska}
\email{tuchol@tiger.chem.uw.edu.pl}
\author{Marcin Modrzejewski}
\author{Robert Moszynski}

\affiliation{Faculty of Chemistry, University of Warsaw, Pasteura 
1, 02-093 Warsaw, Poland}

\begin{abstract}
Theory of one-electron transition density matrices has been 
formulated within the time-independent coupled cluster method
for the polarization propagator [R. Moszynski, P. S. Żuchowski, 
and B. Jeziorski, Coll.
Czech. Chem. Commun. {\bf 70}, 1109 (2005)].
Working expressions have been obtained and implemented with
the coupled cluster method limited to
single, double, and linear triple excitations (CC3).
Selected dipole and quadrupole transition probabilities of the
 alkali earth atoms, computed with the new transition
density matrices are compared to the experimental data. 
Good agreement between theory and experiment is found. 
The results obtained with the new approach are of the same 
quality as the results obtained with the linear response coupled cluster theory (LRCC).
The one-electron density matrices for the ground state in the
 CC3 approximation 
have also been implemented. The dipole moments for a few representative diatomic molecules
have been 
computed with several variants of the 
new approach, and the results are discussed to choose the
 approximation with 
the best balance between the accuracy and computational efficiency.
\end{abstract}
\begin{center}
\maketitle
\end{center}

\section{Introduction}
One of the most challenging problems of modern quantum chemistry is an
accurate and fast computation of molecular properties.
Coupled cluster theory (CC) which is the gold standard of 
quantum chemical
methods, combines an accurate 
description of the electronic structure with an affordable
computational cost for medium sized molecules.
The coupled cluster Ansatz is presented as
\cite{coester1958bound, cizek1969use, paldus1975time, bartlett1989coupled, crawford2000introduction, bartlett1995modern, bartlett2007coupled, vcarsky2010recent, lyakh2011multireference}
\equl{\Psi = e^T \Phi,}{In-antz}
where the cluster operator $T$ for an $N$ electron system is the 
sum of single, double, and higher excitations, 
$T=T_1+T_2+\cdots + T_{N}$, and $\Phi$ 
is the reference function.
Due to the exponential form of the Ansatz,
the CC theory  is size-extensive for
any truncation of $T$.
The possibility of restricting $T$
 to a particular excitation level introduces a hierarchy of
approximations: coupled cluster singles and doubles (CCSD), 
coupled cluster singles, doubles, and triples (CCSDT), etc.
Also, the methods CC2 \cite{christiansen1995second} and CC3,\cite{koch1997cc3}
approximating CCSD and CCSDT, respectively, were developed. 
The CC3 equations for $T_1$ and $T_2$ have the same form as in CCSDT.
 The equation for $T_3$, however, includes only terms up to the second order
in the fluctuation potential.
The CC3 approximation ensures that the triple amplitudes
are correct through the second order, 
while there is no need for storing $T_3$ in memory: 
they are readily computable on the fly 
with expressions including
single and double excitations.
The ground state CC3 model scales as $\mathcal{N}^7$, 
whereas CCSDT scales as $\mathcal{N}^8$, with the size 
of the basis $\mathcal{N}$.

Currently, molecular properties of the ground state within the 
CC framework are computed as 
the derivative of the first-order Lagrangian with
respect to the field strength.\cite{halkier1997first, hald2002calculation} 
An alternative method, referred to
as XCC,  was proposed by \citet{jeziorski1993explicitly}
and further investigated by Moszynski et al. \cite{moszynski1993many, moszynski1994many}, \citet{korona2006one} and 
\citet{korona2006time}. In the XCC approach,  the first-order properties
are computed directly from  the definition of the 
quantum-mechanical expectation value. 
This formalism
is conceptually simple and
its  computational cost is lower
than in the case of the Lagrangian technique as it does not 
require finding the 
expensive left-hand solution of the CC equations, 
the so-called $\Lambda$ or $Z$ vector. 
\cite{halkier1997first, hald2002calculation}

The main object of interest in this study is the linear 
response function $\odp$, often referred to in the literature 
as the polarization propagator. The linear response function
 describes the response of an observable
 $X$ to the perturbation $Y$ oscillating with the frequency 
$\omega$. The residues of the polarization
propagator are connected to many physical observables, 
e.g. transition probabilities, lifetimes, and line strengths.
For real $\omega$ and for purely real or purely 
imaginary perturbations $Y$, the polarization propagator 
satisfies the following relation
\equl{\odp =\odpm,}{In-trs}
which reflects the time-reversal symmetry.

The linear response function within  CC theory can be 
computed either from
the response theory (LRCC),
\cite{koch1990coupled, koch1994caclculation, koch1997coupled}
 or from the time-independent XCC theory.
\cite{moszynski2005time}
Both theories give the polarization propagator
 satisfying \fr{In-trs}.
In the LRCC approach the 
time-reversal symmetry of the 
linear response function 
follows from the
restriction of the time-dependent expectation value to the real part, 
which is  otherwise not guaranteed to be real if an 
approximate coupled cluster wave function is employed.
In XCC, one starts from the exact expression for
the polarization propagator. Thus, the correct symmetry 
is present in the XCC theory
from the start. The final form of the polarization 
propagator in this theory
is Hermitian in the sense that any truncation of the 
cluster operators does not
violate the correct time-reversal symmetry.


During the twenty years since 
the initial formulation of the XCC method,\cite{jeziorski1993explicitly} 
numerous studies restricted to the CCSD level were reported: 
electrostatic\cite{moszynski1993many} and exchange\cite{moszynski1994many}
 contributions to the interaction energies
of closed-shell systems,
first-order molecular properties,\cite{korona2006one} 
static and dynamic dipole polarizabilites,
\cite{korona2006time} frequency-dependent 
density susceptibilities employed in SAPT(CC).\cite{korona2008dispersion} 
In this paper we present the derivation and implementation 
of the transition density matrices obtained from the XCC linear response 
function \cite{moszynski2005time} at the CC3 level. 
Also, the results for the first-order one-electron properties at the CC3 level
are presented in order to test various approximations to 
the XCC theory.

This paper is organized as follows.
In \frs{subseco-th} we 
 derive the formula for the first-order properties
within the XCC3 theory.
We also report the derivation of the
transition density matrices from the XCC linear response function.
Next, in \frs{numres} we present the numerical results for the ground-state dipole
moments of some representative diatomic molecules. 
We discuss various  approximations to 
the XCC3 theory
that offer the best balance between the accuracy and 
computational efficiency.
We continue the discussion of the results with the atomic dipole and quadrupole 
 transition probabilities computed within the XCC3 theory.
Whenever possible, extensive comparison with the 
experimental data as well as with the data obtained from the LRCC3 calculations
is reported. Finally in \frs{concl} we conclude our paper.

\section{Theory}\label{subseco-th}
\subsection{Basic definitions}
All the operators in this work are expressed through the 
singlet orbital replacement operators\cite{paldus1988clifford}
\equl{E_{pq} = a_{p\alpha}^\dagger a_{q\alpha} + a_{p\beta}^\dagger a_{q\beta},}{ugg}
which satisfy the commutation relation 
$[E_{pq}, E_{rs}] = E_{ps}\delta_{rq} - E_{rq}\delta_{ps}$. From now on, 
$a, b, c \ldots$ and $i, j, k \ldots$ denote virtual
 and occupied orbital indices, 
 respectively,
and $p, q, r \ldots$ general indices.
The cluster operator $T$ is represented in a compact form
as a sum of $n$-tuple excitation operators $T_n$,
\equ{
{T}_n = \frac{1}{n!}\sum_{\mu_n} t_{\mu_n} {\mu_n},
}
where $\mu_n$ stands for the
product of the $n$ singlet excitation operators $E_{ai}E_{bj}\cdots E_{fm}$.
The CC amplitudes  satisfy the following  
 permutation symmetry relations
\equs{&t_{ij}^{ab} = t_{ji}^{ba} \\
&t_{ijk}^{abc} = t_{ikj}^{acb}=t_{jik}^{bac}=t_{jki}^{bca}=t_{kij}^{cab}=t_{kji}^{cba}.}

The excitation energies in this work are obtained from the diagonalization of the 
CC Jacobian matrix,\cite{sekino1984linear, koch1990coupled, helgaker2013molecular} 
\equl{A_{\mu_n\mu_m} = \braket{\widetilde{\mu}_n}{\cm{e^{-T}He^T}{{\mu_m}}},
}{form-jacob}
where we introduce the shorthand notation 
$\braket{X}{Y} = \braket{X\Phi}{Y\Phi}$, $\brz{X}=\braket{\Phi}{X\Phi}$. 
The elements of the Jacobian are defined in the 
biorthonormal basis 
\equ{\langle \widetilde{\mu}_n|\nu_n\rangle = \delta_{\mu_n\nu_n}
}
For the single and double excitation manifold we used the basis 
proposed by \citet{helgaker2013molecular}.
A biorthonormal
and nonredundant basis for the triply excited manifold is derived in the Appendix.

The expectation value of an observable in the XCC theory is given 
by the explicitly connected, size-consistent  expression introduced
 by \citet{jeziorski1993explicitly}
\equl{\bar{X}=\langle e^{S^\dagger}e^{-T}Xe^{T}e^{-S^\dagger}\rangle.
}{In-av}
The auxiliary operator $S = S_1 + S_2 + \cdots + S_N$ is the 
solution of the following equation
\equsl{
S_n &= T_n  - \frac1n \hat{\mathcal{P}}_n \left ( \sum_{k=1}
\frac{1}{k!}\comute{\widetilde{T}\dgg}{T}_k \right ) \\
&- \frac1n \hat{\mathcal{P}}_n\left (\sum_{k=1}\sum_{m=0}
\frac{1}{k!}\frac{1}{m!} [\comute{\widetilde{S}}{\dg{T}}_k,T]_m\right),
}{ss}
where \equ{
\widetilde{T} = \sum_{n = 1}^{N} nT_n, \qquad\widetilde{S} = \sum_{n = 1}^{N} nS_n,
}
and
\equ{ [A, B]_k = \underbrace{[[\cdots[[A, B], B] \cdots ]}_{\textrm{nested } k \textrm{ times}}.
}
The superoperator $\hat{\mathcal{P}}_n(X)$ 
projects the $n$-tuple
excitation part of an arbitrary
 operator $X$
\equ{\hat{\mathcal{P}}_n(X) = \frac{1}{n!}\sum_{\mu_n} \braket{\widetilde{\mu}_n}{X}{\mu_n}.
}

The expanded expression for $S_n$, \fr{ss}, is finite, though it 
contains cumbersome terms with multiply-nested commutators.
These terms are of high order in the 
fluctuation potential.\cite{jeziorski1993explicitly} 
Also, the r.h.s. of \fr{ss} depends on $S$, therefore solving this 
equation requires an iterative procedure. 
However, $S$ can efficiently be approximated while retaining 
the size consistency 
of the expectation value expression. 
Below, we present the expressions for $S_n(m)$ 
for $n\in \{1,2,3\}$ and $m\in\{2,3,4\}$, with $m$ denoting the 
highest many-body perturbation theory (MBPT) order fully included,

\equsl{
S_1(2) &= T_1   \\
S_1(3) &= S_1\dr +  \hat{\mathcal{P}}_1\left ([T_1^{\dagger}, T_2] \right )  \\
&+ \hat{\mathcal{P}}_1\left ([T_2^{\dagger}, T_3] \right )   \\
S_1(4) &= S_1\tr +  
\hat{\mathcal{P}}_1\left ([[T_2^{\dagger}, T_1], T_2] \right )\\
&+\frac12\hat{\mathcal{P}}_1\left ([[T_3^{\dagger}, T_2], T_2 ]\right )  \\
S_2(2) &= T_2  \\
S_2(3) &= S_2\dr + \frac12\hat{\mathcal{P}}_2\left ([[T_2^{\dagger}, T_2], T_2] \right )   \\
S_2(4) &=  S_2\tr  + \hat{\mathcal{P}}_2\left ([T_1^{\dagger}, T_3] \right )   \\
S_3(2) &= T_3   \\
S _3(4) &=  S_3\tr + \frac12\hat{\mathcal{P}}_3\left ([[T_1^{\dagger}, T_2], T_2] \right )  \\
&+ \hat{\mathcal{P}}_3\left ([[T_2^{\dagger}, T_2], T_3] \right ) \\
}{przybl-s}

We test the accuracy of three approximations denoted as XCC3S($m$), with $m=2,3,4$
\equsl{
&\mathrm{XCC3S(2)}: S_1(2) + S_2(2) + S_3(2)\\
&\mathrm{XCC3S(3)}: S_1(3) + S_2(3) + S_3(2)\\
&\mathrm{XCC3S(4)}: S_1(4) + S_2(4) + S_3(2).
}{xcc3s}
One should note that in all three approximations $S_3 = T_3$.

The accuracy of $S$ depends on the underlying wave function model.
The CC3 method includes
$T_1$ and $T_2$ correct through \bigoh{third} and $T_3$ 
correct through \bigoh{second}.
The accuracy of $S_1$, $S_2$, and $S_3$ is of the same 
order of MBPT as the accuracy of the corresponding
$T_1$, $T_2$, and $T_3$ amplitudes.
The lowest order contributions to $S_4$ are of the third order, but
this quantity appears only in \bigoh{fourth} contributions 
to the transition density matrices,
and is not required.

Using the commutator expansion in \fr{In-av}
we obtain the following 
formula for the expectation value of an operator at the CC3 level of theory
\equsl{\bar{X} &=  \sum_{\mathcal{M}=0}^8 \bar{X}^{(\mathcal{M})}=  \brz{X}^{(0)} \\ 
&+ \braket{S_1}{X}^{(2)} + \brz{[X, T_1]}^{(2)} + \braket{S_2}{[X, T_2]}^{(2)} \\ 
&+ \braket{S_1}{[X, T_2]}^{(3)} + \braket{S_2}{[X, T_3]}^{(3)}\\ 
&+ \braket{S_1}{[X, T_1]}^{(4)} + \braket{S_2}{[[X, T_1], T_2]}^{(4)} \\ 
&+ \braket{S_3}{[X, T_3]}^{(4)} + \frac12\braket{S_3}{[[X, T_2], T_2]}^{(4)}\\ 
&+\frac12\braket{S_1^2}{[X, T_2]}^{(5)}+\frac12\braket{S_1S_2}{[[X, T_2], T_2]}^{(5)}  \\ 
&+\frac12\braket{S_1S_2}{[X, T_3]}^{(5)}\\ %
&+\frac12\braket{S_1}{[[X, T_1], T_1]}^{(6)}+\frac12\braket{S_1^2}{[X, T_3]}^{(6)}\\ 
&+\frac12\braket{S_1^2}{[[X, T_1], T_2]}^{(7)}\\ 
&+\frac{1}{12}\braket{S_1^3}{[[X, T_2], T_2]}^{(8)}+\frac16\braket{S_1^3}{[X, T_3]}^{(8)}. 
}{Th-Fo-av}
The upper index of $\bar{X}^{(\mathcal{M})}$ indicates
 an $\mathcal{M}$-th order contribution.
Apart from  $T_n$ and $S_n$ for $n>3$, no other
 approximations have been introduced in \fr{Th-Fo-av}.

\subsection{XCC3 transition density matrices}\label{sec-trans}
In the exact theory the polarization propagator is defined 
by the following expression\cite{oddershede1987propagator}
\equsl{\odp = &-\braket{\Psi_0}{Y\frac{Q}{H - E_0 +\omega}X\Psi_0}\\
&-\braket{\Psi_0}{X\frac{Q}{H - E_0-\omega}Y \Psi_0},
}{odp-exact}
where $H$ denotes the Hamiltonian, 
$\Psi_0$ is the normalized ground-state wave function, 
$E_0$ is the ground state energy,  and $Q$ is the 
projection operator on the 
space spanned by all excited states.
The line strength   $S^{0K}_{XY}$ of the transition 
to the $K$-th excited state 
is obtained as the residue of the 
linear response function:
\equm{
\lim_{\omega \rar \omega_K}(\omega - \omega_K)\odp
= \\\sum_{K'}\braket{\Psi_0}{X \Psi_{K'}}\braket{\Psi_{K'}}{Y\Psi_0}
 = S_{XY}^{0K}
}
\equm{\lim_{\omega \rar -\omega_K}(\omega + \omega_K)\odp
= \\-\sum_{K'}\braket{\Psi_0}{Y \Psi_{K'}}\braket{\Psi_{K'}}{X\Psi_0}
 =  S_{XY}^{K0}
}
where $K'$ runs over all degenerate states corresponding 
to the excitation energy $\omega_K$.
The time-reversal symmetry, \fr{In-trs}, is transferred 
from the polarization propagator
to the line strength $ S_{XY}$ through the relation
\equl{S_{XY}^{0K} = -(S_{XY}^{K0})^{\star} .
}{In-ss}
\citet{moszynski2005time} have expressed the polarization 
propagator within the 
framework of the XCC theory
\equml{
\langle\langle X;Y \rangle \rangle_\omega = \\
\braket{e^{-S}e^{T^\dagger}Ye^{-T^\dagger}e^{S}}{\hat{\mathcal{P}}\left(e^{S^\dagger}\Omx e^{-S^\dagger}\right)}
 + \mbox{g.c.c.},}{In-odp}
where g.c.c. (generalized complex conjugate) denotes the 
complex conjugation of the r.h.s. and substitution 
of $\omega$ for 
$-\omega$.
Not only this expression satisfies the time 
reversal symmetry, but is also size-consistent because
 it can solely be represented in terms of commutators.

The operator $\Ome{}{X}$ appearing in \fr{In-odp} is 
solution of 
the  linear response equation,\cite{moszynski2005time} 
\equl{
\braket{\widetilde{\mu}}{\comute{e^{-T} H e^T}{ \Ome{}{X}} + \omega\Ome{}{X}
  + e^{-T}Xe^T} = 0,
}{om}
where $\Ome{}{X} = \Omega_1^X(\omega) + 
\Omega_2^X(\omega) + \cdots + \Omega_N^X(\omega)$,
and  $\Omega_n^X(\omega)$ is an excitation operator of the form
\equ{
\Omega_n^X =
\sump_{\mu_n} O_{\mu_n}^X(\omega) {\mu_n}.
}
where $\sum_{\mu_n}'$ stands for restricted summation over 
non-redundant excitations
for double excitations $ai \ge bj$ and for triple
 excitations $ai \ge bj \ge ck$.
Using the transformation from the molecular orbital 
basis to the Jacobian basis
\equl{
\mu_n = \sum_M \mathcal{L}_{\mu_n M}^{\star}  r_M,\qquad
 \widetilde{\mu}_n^{\star}   = \sum_M \mathcal{R}_{\mu_n M}^{\star} l_M^{\star}  
}{Th-jbas}
$\Omega^X(\omega)$ can be written as
\equsl{
\Omx &= \sum_M \sum_{n=1}^{N} \sump_{\mu_n}
\mathcal{L}_{\mu_n M}^{\star} O_{\mu_n}^X(\omega) r_{M} \\
&= \sum_{M} O_{M}^X(\omega) r_{M}.
}{om-jak}
 \fr{om}
takes then the form
\equml{
\braket{l_M}{\comute{\etm H\et }{r_M}}O_M^X(\omega) \\+
\omega O_M^X(\omega)
  + \braket{l_M}{e^{-T}Xe^T} = 0,
}{jakobian}
where $\braket{l_M}{\comute{\etm H\et }{r_M}}$ 
is the $M$-th excitation energy $\omega_M$, and we used the 
biorthonormality condition $\braket{l_M}{r_K}=\delta_{MK}$.
The $O_M^X(\omega)$ reads
\equl{
O_M^X(\omega) = -\frac{\braket{l_M}{e^{-T}Xe^T}}{\omega_M+\omega}.
}{om-jac}

We will now translate \fr{In-odp} into a computationally 
transparent form.
The action of the projection superoperator 
$\hat{\mathcal{P}} = \hat{\mathcal{P}}_1 + 
\hat{\mathcal{P}}_2 + \cdots + \hat{\mathcal{P}}_N$ 
on the commutator expansion of $e^{S^\dagger}\Omx e^{-S^\dagger}$
produces a sum  of multiply nested commutators
\equal{&\hat{\mathcal{P}}\left(\sum_{n=1}^{N}
  \sump_{\mu_n}\sum_{k=0}^{n-1}
  \frac{1}{k!}[S^\dagger,O_{\mu_n}^X(\omega)]_k\right) = \\
&\sum_{n=1}^{N}
  \sump_{\mu_n}O^X_{\mu_n}\sum_{k=0}^{n-1}\frac{1}{k!}\sum_\Gamma
    [\dg{S_{n_1}},[\cdots[\dg{S_{n_{k-1}}},[\dg{S_{n_{k}}},\mu_n]\cdots]],\nonumber}{pombch}
where the last summation runs over all sequences satisfying the condition 
\equ{\Gamma: k \leq n_1 + \cdots + n_k \leq n-1.}
Using \fr{pombch}, 
the polarization propagator in  the molecular orbital basis takes the form
\equl{ \odp = \sum_{n=1}^{N}\sump_{\mu_n}O^X_{\mu_n} 
(\omega)\gamma_{\mu_n}^Y + \mbox{g.c.c.,}}{resp-mo}
where we use the shorthand notation for 
$\gamma_{\mu_n}^Y$ and $\eta({\mu_n})$ respectively
\equs{&\gamma_{\mu_n}^Y =
  \brz{\esd\etm Y\et\esdm\eta(\mu_n)},\\
&\eta({\mu_n}) = \sum_{k=0}^{n-1}\frac{1}{k!}\sum_\Gamma
 [\dg{S_{n_1}},[\cdots[\dg{S_{n_{k-1}}},[\dg{S_{n_{k}}},\mu_n]\cdots]].}
Transformation of \fr{resp-mo} to the Jacobian basis 
leads to the following expression
\equal{
& \ \langle\langle X;Y\rangle\rangle_\omega  = \nonumber\\
&=-\sum_M\frac{\big\langle l_M \big| e^{-T}Xe^T \big\rangle  \big\langle\esd\etm Y\et\esdm\eta(r_M)\big\rangle}{\omega_M+\omega} + \mbox{g.c.c.,}\nonumber\\
&= -\sum_M\frac{\xi^X_{M}\gamma_M^Y}{\omega_M+\omega} + \mbox{g.c.c.,}
}{odp-jak}
where
\equ{
\begin{split}
&\begin{split}
\xi_M^X &= \big\langle l_M \big| e^{-T}Xe^T \big\rangle\\
&=\sum_{n=1}^{N}\sump_{\mu_n}\mathcal{L}_{\mu_n M}\braket{{\mu_n}}{e^{-T}Xe^T} \\
&=  \sum_{n=1}^{N}\sump_{\mu_n}\mathcal{L}_{\mu_n M}\xi_{\mu_n}^X.
\end{split}\\
&\begin{split}
\gamma_M^Y & = \big\langle\esd\etm Y\et\esdm\eta(r_M)\big\rangle\\
& = \sum_{n=1}^{N}\sump_{\mu_n}\mathcal{R}_{\mu_n M}  
\brz{e^{S^\dagger}e^{-T}Ye^{T}e^{-S^\dagger}\eta(\mu_n)}\\ 
& = \sum_{n=1}^{N}\sump_{\mu_n}\mathcal{R}_{\mu_n M}\gamma^Y_{\mu_n}.
\end{split}
\end{split}
}

The transition strength matrices are computed as the
residues of the XCC linear response function
\equ{S_{XY}^{0K} = 
 -\sum_{K'}\gamma_{K'}^Y
\xi_{K'}^X  \qquad
S_{XY}^{K0}
= \sum_{K'}(\gamma_{K'}^Y)^{\star} 
(\xi_{K'}^X)^{\star}  . 
}
The line strengths are connected 
by the relation of antihermiticity, \fr{In-ss}, which
 comes up naturally 
in the XCC formalism.
As our formulas for the transition strength matrices
are exclusively expressed in terms of commutators, 
they are automatically size intensive, regardless
of any truncation of the $T$ or  $S$ operators.

We now present 
the scheme of approximations to the 
product 
\equl{\gamma_{K}^Y \xi_{K}^X = \sum_{n=1}^{N}\sump_{\mu_n}\mathcal{R}_{\mu_n M} 
\gamma^Y_{\mu_n}  \sum_{m=1}^{N}  \sump_{\mu_m}\mathcal{L}_{\mu_m M}\xi_{\mu_m}^X
.}{xiga}

The explicit expressions for $\gamma_{\mu}^Y$ and $\xi_{\mu}^X$ 
 in the CC3 approximation
are:
\equl{
\begin{split}
&\begin{split}
(\gamma^Y_{\mu_1})^{\mathrm{CC3}} &= \langle  (Y + \cm{S_1^\dagger}{Y}  + 
\cm{\dg{S_2}}{Y} +
    \cmmb{\dg{S_2}}{Y}{T_1}\\&
+\cmmb{\dg{S_2}}{Y}{T_2} +\cmmb{\dg{S_3}}{Y}{T_2}) \mu_1\rangle,
\end{split}\\
&(\xi_{\mu_1}^X)^{\mathrm{CC3}}= \langle  {\mu_1}| X + \cm{X}{T_1} + \cm{X}{T_2} \rangle,\\
&\begin{split}
(\gamma^Y_{\mu_2} )^{\mathrm{CC3}} &=  \langle  (\cm{S_2^\dagger}{Y} + \cm{S_3^\dagger}{Y} + \cm{S_2^\dagger}{\cm{S_1^\dagger}{Y}}\\
&+\cm{S_2^\dagger}{\cm{Y}{T_1}} + \cm{S_3^\dagger}{\cm{Y}{T_2}} ) \mu_2\rangle \\
&+ \langle (Y + \cm{S_2^\dagger}{Y})\cm{\dg{S_1}}{{\mu_2}}\rangle,\\
\end{split}\\
&\begin{split}
(\xi_{\mu_2}^X)^{\mathrm{CC3}} &= \langle
   {\mu_2}| \cm{X}{T_2}+ \cm{X}{T_3}\\
&+  \cmm{X}{T_1}{T_2} \rangle,\\
\end{split}\\
&\begin{split}
(\gamma^Y_{\mu_3})^{\mathrm{CC3}} &= \langle (\cm{S_3^\dagger}{Y} + \cm{S_2^\dagger}{\cm{S_1^\dagger}{Y}} \\
&+ \frac12\cm{S_2^\dagger}{\cm{S_2^\dagger}{Y}})\mu_3\rangle + \langle \cm{S_2^\dagger}{Y}\cm{\dg{S_1}}{{\mu_3}}\rangle, \\
&+ \langle (Y + \cm{S_1^\dagger}{Y}+ \cm{S_2^\dagger}{Y})\cm{\dg{S_2}}{{\mu_3}}\rangle, \\
\end{split}\\
&\begin{split}
(\xi_{\mu_3}^X )^{\mathrm{CC3}} &= \langle \mu_3| \cm{X}{T_3} + \frac12\cmm{X}{T_2}{T_2}\\
&+ \cmm{X}{T_1}{T_2} \rangle.
\end{split}
\end{split}
}{n2}
The expressions for $\gamma_{\mu}^Y$ and $\xi_{\mu}^X$ contain contributions
up to and including the third order of MBPT. In $\gamma_{\mu_2}^Y$
and $\gamma_{\mu_3}^Y$ we have omitted the third order terms
$\frac12\langle [S^{\dagger}_2, [S^{\dagger}_2, [Y, T_2]]]\mu_2\rangle$
and
$\frac12\langle [S^{\dagger}_2, [S^{\dagger}_2, [Y, T_2]]]\mu_3\rangle$  as they are  computationally  much more
demanding than the rest of the contributions.
The $S_1$ and $S_2$ operators are correct through \bigoh{third}, and
the $S_3$ operator contains only the leading term
correct through \bigoh{second}, \fr{przybl-s}. 

All the implementation-ready formulas presented in this work 
have been derived
with the assistance of the \verb+Paldus+ program developed 
in our laboratory. \verb+Paldus+ is a program for an automated 
implementation of any level of theory
expressible through the products of singlet orbital replacement operators. 
The formulas obtained with \verb+Paldus+ program are 
automatically optimized and incorporated into 
the  parallelized, 
standalone  \verb+Fortran+ code.

\section{Numerical results and discussion}\label{numres}
\subsection{First-order properties at the CC3 level of theory}
\label{subseco-num}
We present the results for the ground-state dipole moments of 
diatomic molecules calculated at the XCC3 level of theory.
The geometries of the diatomic molecules are kept at their 
equilibrium values.\cite{nist} Comparison is done with the
experimental data\cite{lide2010crchandbook} and with the LRCC3 results.
For all  the molecules we  employ the def2-QZVPP basis set.\cite{weigend2005balanced}

\Frff{hf}{cs} 
show the unsigned percentage error  of the dipole moment 
relative to the experimental value {\errname} $ := $ {\relerror} 
as a function 
 of the highest-order term included in \fr{Th-Fo-av}.
In each plot, separate lines represent approximations 
to the auxiliary operator $S$, denoted as XCC3S($m$).
Thus, there are two levels of approximation: one for 
the expectation value formula, 
\fr{Th-Fo-av}, and one for the operator $S$, \fr{xcc3s}.

In each case, the convergence of the expectation value 
defined by \fr{Th-Fo-av}
is achieved after including the terms up to and including 
\bigoh{fifth}.
However, the inclusion of the higher-order terms does not 
introduce much additional computational
costs. 
The most time consuming terms that scale as $\mathcal{N}^8$ 
appear in the fourth and higher orders.
Introduction of intermediates reduces the scaling of all 
such terms to $\mathcal{N}^7$.
As the most expensive terms appear already in the fourth 
order, computing the full sum, \fr{Th-Fo-av},
is essentially of the same cost as computing only the partial sums.

An inspection of  \Frff{hf}{cs} shows that 
in all three cases
the use of XCC3S(3) brings an improvement over XCC3S(2) relative to the
experimental values.
The most challenging case is the CO molecule. For this system, 
the XCC3S(2) level
of theory is unacceptable  with {\errname} reaching $90\%$.
A huge reduction of this error is observed for XCC3S(3) and XCC3S(4).

Importantly, in every case 
 improving the accuracy
of $S$ improves the accuracy of the results. However, 
going from XCC3S(3) to XCC3S(4) brings only a negligible
improvement not worth the corresponding increase in the 
computational complexity, from
$\mathcal{N}^7$ to $\mathcal{N}^8$. We thus recommend 
the XCCS(3) level of theory; this
will be the approximation of $S$ employed to compute 
second order properties.

\begin{figure}[!ht]
\includegraphics[width=\columnwidth]{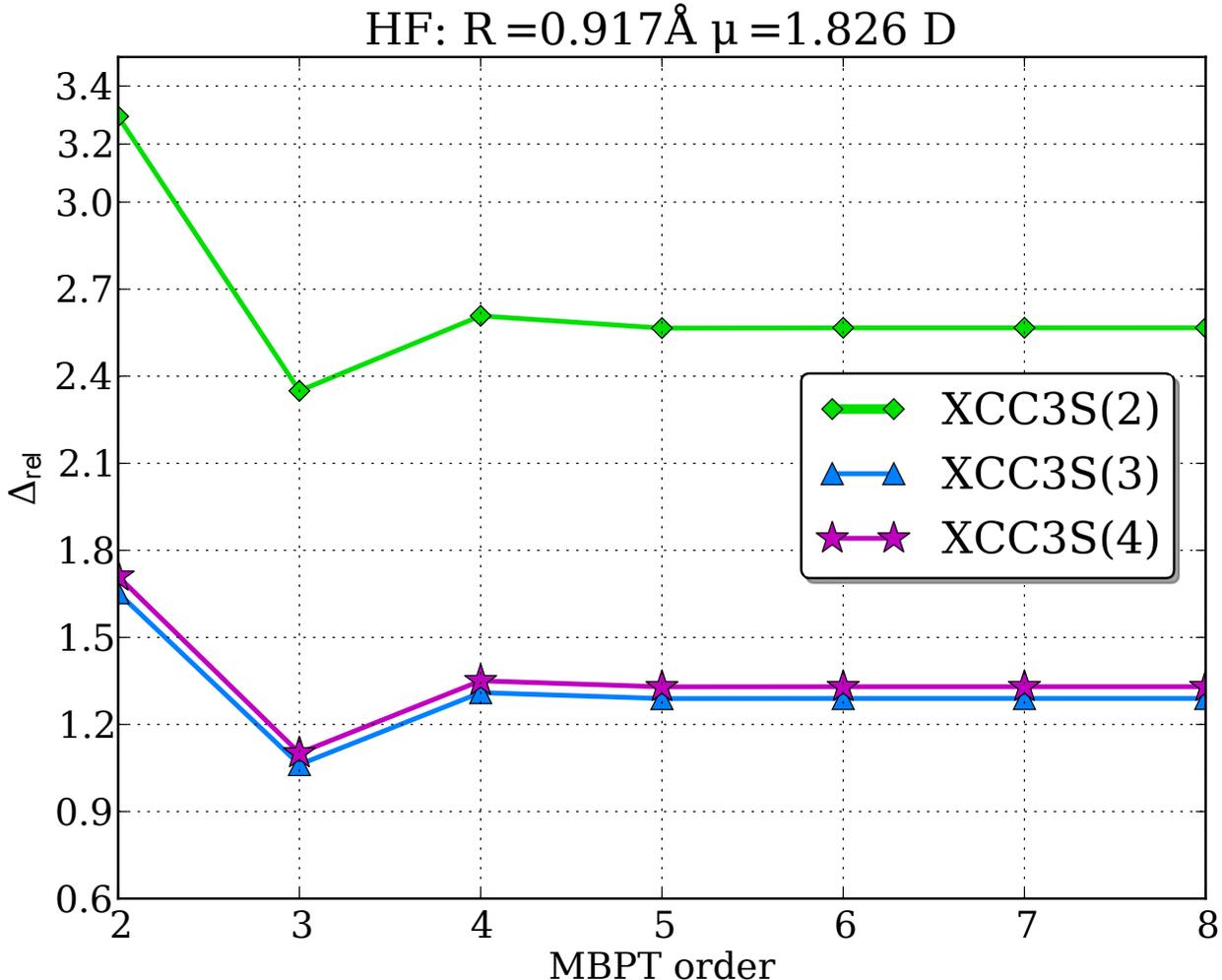}
\caption{{\errname}    of the dipole moment of HF.}
\label{hf}
\end{figure}
\begin{figure}
\includegraphics[width=\columnwidth]{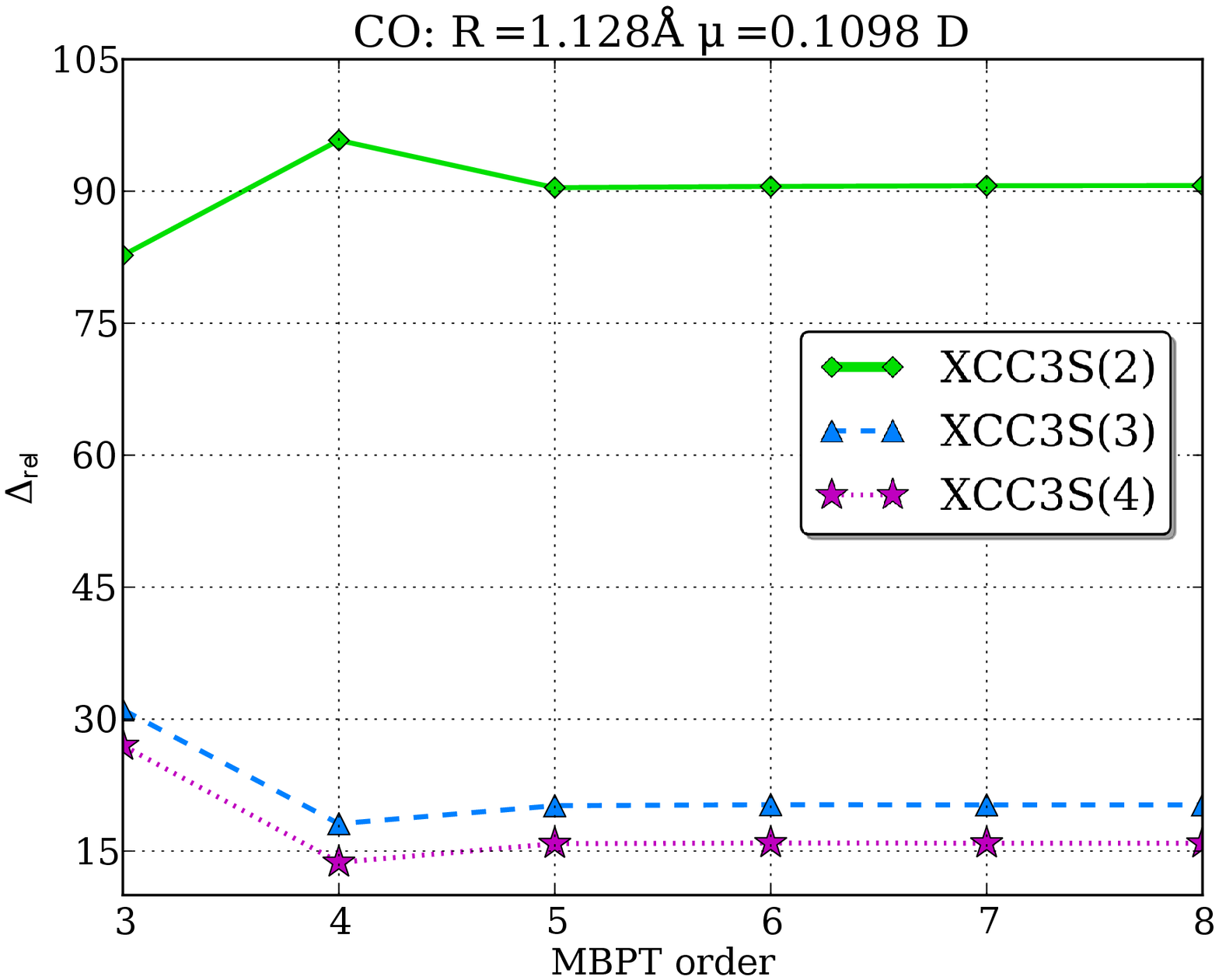}
\caption{{\errname}  of the dipole moment of CO.}
\label{co}
\end{figure}
\begin{figure}[!ht]
\includegraphics[width=\columnwidth]{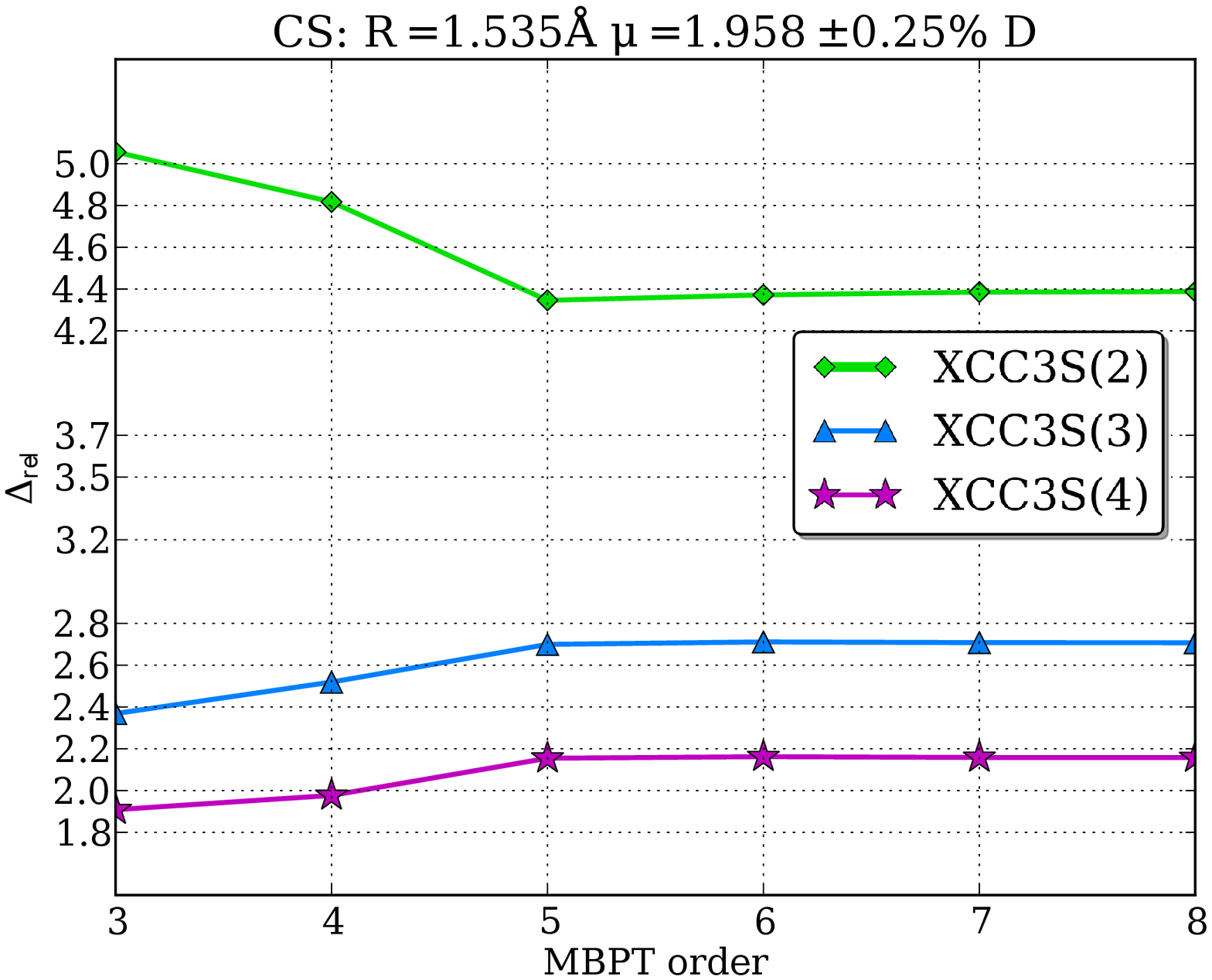}
\caption{{\errname}  of the dipole moment of CS.}
\label{cs}
\end{figure}

\begin{table}[!ht]
\caption{Dipole moments computed with the XCC3S(3) and LRCC3 methods. 
The def2-QZVPP basis set was employed for molecules at 
 equilibrium geometries. The experimental data are given in Debye, 
and the
computed values are given as an signed error 
$\Delta_{\mathrm{method}} = \mu_{\mathrm{exp}} - \mu_{\mathrm{method}}$.}
\label{tabela-cc3-lrcc}
\begin{center}
  \begin{tabular}{l d d d}
		\hline
                \hline
\multicolumn{1}{c}{molecule} & \multicolumn{1}{c}{exp.} & 
\multicolumn{1}{c}{$\Delta_{\mathrm{XCC3S(3)}}$} & \multicolumn{1}{c}{$\Delta_{\mathrm{LRCC3}}$}\\
		\hline
LiH	&	5.884	&	0.0400	&	0.0463\\
HF	&	1.826	&	0.0235	&	0.0071\\
LiF	&	6.3274	&	0.0179	&	0.0879\\
CO	&	0.1098	&	0.0222	&	-0.0264\\
NaLi	&	0.463	&	-0.0107	&	-0.0263\\
HCl	&	1.1086	&	0.0169	&	-0.0216\\
NaF	&	8.156	&	-0.0015	&	0.0812\\
CS	&	1.958	&	0.0530	&	0.0055\\
                \hline
	\end{tabular}
\end{center}
\end{table}

We compare our method with the Lagrangian technique of Hald and J{\o}rgensen.\cite{hald2002calculation} 
\Frt{tabela-cc3-lrcc} shows the signed absolute errors of both 
methods applied to the dipole
moments of the test set of diatomics with the experimental data.
On the average the XCC3S(3) method is only slightly better than LRCC3. 
Indeed, the mean absolute error for XCC3S(2) is equal to 0.023 
and for LRCC3 is equal to 0.038.

This result is encouraging since the XCC3 method is conceptually 
simpler and computationally less demanding than the LRCC3 approach. 
While both methods employ the same model for the ground-state 
wave function (that scales as $v^4o^3$, where $v$ and $o$ 
stand for the number of the virtual and occupied orbitals, respectively), 
the difference lies in the computation of the auxiliary operators required 
for the one-electron properties, i.e. the Lagrangian multipliers 
in the case of the LRCC approach and the operator S in the case 
of the XCC method. The equations for the singles and doubles 
Lagrangian multipliers are solved iteratively and each iteration scales like $v^4o^2$, 
whereas the amplitudes of the $S_1$ and $S_2$ operators are 
computed directly in a single step that scales as $v^3o^3$. 
Moreover, $S_3$ can efficiently be approximated by $T_3$, 
whereas the most expensive, triples Lagrange multipliers in the 
LRCC3 approach have to be computed separately. 
The computational complexity of assembling the density matrices 
from the auxiliary amplitudes, ground-state amplitudes, 
and molecular integrals is the same in both approaches and scales as $v^4o^3$.

\subsection{Transition probabilities} \label{subsect-num}
We have performed computations of the electric
dipole transition probabilities between the \ce{^1S} and \ce{^1P} states 
for the Mg, Ca, Sr and Ba atoms, 
and of the quadrupole transition probabilities 
between the \ce{^1S} and \ce{^1D} 
states  for the Ca and Ba atoms. 

The line strength of the dipole transition  is defined as
\equ{S_d = \sum_{K,K'}| \langle K|\mathbf{d}|K'\rangle |^2}
where $K$ and $K'$ run over all degenerate states, and $\mathbf{d}$ 
is the dipole moment operator.
The dipole transition probability $A_{\ce{^1P ^1S}}$ is related to
the line strength by the relation\cite{drake2006springer} 
\equl{A_{\ce{^1P ^1S}} = \frac{1}{3} \frac{16\pi^3}{3h\epsilon_0\lambda^3}S_d^{\ce{^1P ^1S}},
}{Dt-spring}
where SI units are used for  
$A_{\ce{^1P ^1S}}$, $S_d$ and $\lambda$: s$^{-1}$, m$^2$ C$^2$ and m respectively.

The strength 
of a quadrupole transition is defined as\cite{shortley1940computation}
\equ{S_q = \sum_{K,K'}|\langle K|\mathbf{Q}|K'\rangle|^2,
}
where $\mathbf{Q}$ is the traceless quadrupole moment operator in the 
Shortley's convention,\cite{shortley1940computation} and the transition 
probability reads 
\equl{A_{\ce{^1D ^1S}} = \frac{1}{5}\frac{32\pi^6}{5h\lambda^5}S_q^{\ce{^1D ^1S}},
}{Qt-short}
where SI units are used for  $A_{\ce{^1D ^1S}}$, $S_q$ and $\lambda$: s$^{-1}$, 
m$^4$C$^2$ and m respectively. $A_{ki}$ will be used as a shorthand notation for 
both dipole and quadrupole transition probabilities.

\subsubsection{Dipole transition probabilities}
\Frt{Dip} shows the atomic transition probabilities  $A_{ki}$ for the \ce{^1S} -\ce{^1P}
transitions in Mg, Ca, Sr, and Ba atoms. The results 
are compared with the available spectroscopic data. In each case 
we performed calculations with the XCC3S(2), XCC3S(3), and LRCC3 methods. 
To illustrate the convergence of the computed dipole
transition probabilities with the basis set size, we use 
a progression of basis sets. 

We also performed computations with the multireference configuration 
interaction (MRCI) method restricted to single and double excitations 
in order to compare our method with approaches based on different 
models of the wave function. Numerical results for the dipole 
transition probabilities are presented in the last two columns of 
\Frt{Dip}. The MRCI results were obtained with the Molpro program.\cite{MOLPRO} 
In all cases, except for the Ba atom, the agreement with the experiment 
of the MRCI data is by an order of magnitude worse than of the 
results obtained with the XCC and LRCC methods.

Except for the Ba case, the results converge quickly 
to the experimental benchmarks 
with the increase of the basis set size.
In all other cases,
for the largest bases employed, the results are 
well within the experimental error bars.
For the Ba atom no improvement of the XCC, LRCC, or the 
MRCI values is observed with the enlargement of the basis. 
This can probably be attributed to the use of the pseudopotential 
that treats the core-electron correlation in an approximate way.
 In the case of Mg, Ca, and Sr atoms the use
of XCC3S(3) shows a significant improvement over XCC3S(2). 
This
corroborates the choice of XCC3S(3) as the recommended approach.
The comparison of XCC3S(3) with LRCC3 shows that the 
transition probabilities  are of the same quality.

Although the transition probabilities obtained with the 
XCC3 and LRCC3 methods are of equivalent quality, the computational 
steps required to obtain these properties differ, with XCC3 being 
the simplest approach. From the computational point of view, the 
major additional cost of LRCC3 is the calculation of the matrix 
$F_{\mu\nu}^X= \brakett{\Lambda}{[[X, \mu] \nu]}{\Psi}$
and obtaining the $\mathbf{F}$-transformed
vectors.\cite{koch1990coupled, koch1997coupled, christiansen19998integral}
Moreover, the LRCC3 approach involves (as in the case of 
ground-state properties) an iterative computation of the 
Lagrange multipliers, while the XCC3 method requires only a 
single step calculation of the $S$ amplitudes. The remaining 
steps, i.e. the diagonalization of the Jacobian matrix and 
solution of the response equation \fr{om}, are the same for both methods.

\begin{table}[!ht]
\caption{Dipole transition probabilities obtained with the XCC3, LRCC3, and MRCI methods. All $A_{ki}$ 
values given in $10^8 s^{-1}$. $\Delta = A_{ki}^{\mt{exp}} - A_{ki}^{\mt{comp}}$. 
T = def2-TZVP\cite{weigend2005balanced}, Q = def2-QZVP\cite{weigend2005balanced},
 5 = cc-pV5Z\cite{peterson2002accurate, koput2002ab}, 
E46 = ECP46MDF\cite{lim2006relativistic}.
} \label{Dip}
\begin{center}
	\begin{tabular}{l d d d d d d d d}
		\hline
                \hline
\multicolumn{9}{c}{\multirow{2}{*}{{\bf Mg} $3s^2-3s3p$: $A_{ki}^{\mt{exp}} = 4.95(15)$ 
\cite{lide2010crchandbook,sansonetti2005handbook}}}\\
&&&&&&&&\\
		\hline
                \multicolumn{1}{c}{\multirow{2}{*}{}} & 
\multicolumn{1}{r}{\multirow{2}{*}{$A_{ki}^{\mt{S(2)}}$ }}
 & \multicolumn{1}{r}{\multirow{2}{*}{$\Delta^{\mt{S(2)}}$}} & 
\multicolumn{1}{r}{\multirow{2}{*}{$A_{ki}^{\mt{S(3)}}$}}
& \multicolumn{1}{r}{\multirow{2}{*}{$\Delta^{\mt{S(3)}}$ }}& 
\multicolumn{1}{r}{\multirow{2}{*}{$A_{ki}^{\mt{LR}}$}}
& \multicolumn{1}{r}{\multirow{2}{*}{$\Delta^{\mt{LR}}$ }} &
\multicolumn{1}{r}{\multirow{2}{*}{$A_{ki}^{\mt{MR}}$}} &
\multicolumn{1}{r}{\multirow{2}{*}{$\Delta^{\mt{MR}}$ }} \\
&&&&&&&&\\
\hline

T&5.808&	-0.858&	5.876&	-0.926&5.882&	-0.932  & 6.04&1.09\\
Q&4.777&	0.173&	4.833&	0.117&4.843&	0.107  & 4.80&-0.15\\
5&4.796&	0.154&	4.853&	0.097&4.864&	0.086 & 4.83&-0.12\\

                \hline
	\end{tabular}

\vspace{10pt}

	\begin{tabular}{l d d d d d d d d}
		\hline
                \hline
\multicolumn{9}{c}{\multirow{2}{*}{{\bf Ca} $4s^2-4s4p$ : $A_{ki}^{\mt{exp}} = 2.20(4)$ 
\cite{morton2003atomic,sansonetti2005handbook}}}\\
&&&&&&&&\\
		\hline
                \multicolumn{1}{c}{\multirow{2}{*}{}} & 
\multicolumn{1}{r}{\multirow{2}{*}{$A_{ki}^{\mt{S(2)}}$ }}
 & \multicolumn{1}{r}{\multirow{2}{*}{$\Delta^{\mt{S(2)}}$}} & 
\multicolumn{1}{r}{\multirow{2}{*}{$A_{ki}^{\mt{S(3)}}$}}
& \multicolumn{1}{r}{\multirow{2}{*}{$\Delta^{\mt{S(3)}}$ }}& 
\multicolumn{1}{r}{\multirow{2}{*}{$A_{ki}^{\mt{LR}}$}}
& \multicolumn{1}{r}{\multirow{2}{*}{$\Delta^{\mt{LR}}$ }} &
\multicolumn{1}{r}{\multirow{2}{*}{$A_{ki}^{\mt{MR}}$}} &
\multicolumn{1}{r}{\multirow{2}{*}{$\Delta^{\mt{MR}}$ }} \\
&&&&&&&&\\
\hline

T&2.352&-0.152&	2.385&	-0.185&2.386& -0.186&2.71&0.51\\
Q&2.183&0.017&	2.211&	-0.011& 2.212& -0.012&2.64&0.44\\
5&2.159  &0.041&	2.184&	0.016& 2.184&0.016&2.62&0.42\\
                \hline
	\end{tabular}

\vspace{10pt}

	\begin{tabular}{l d d d d d d d d}
		\hline
                \hline
\multicolumn{9}{c}{\multirow{2}{*}{{\bf Sr} $5s^2-5s5p$ : $A_{ki}^{\mt{exp}} = 2.01(3)$ 
\cite{sansonetti2010wavelengths,sansonetti2005handbook}}}\\
&&&&&&&&\\
		\hline
                \multicolumn{1}{c}{\multirow{2}{*}{}} & 
\multicolumn{1}{r}{\multirow{2}{*}{$A_{ki}^{\mt{S(2)}}$ }}
 & \multicolumn{1}{r}{\multirow{2}{*}{$\Delta^{\mt{S(2)}}$}} & 
\multicolumn{1}{r}{\multirow{2}{*}{$A_{ki}^{\mt{S(3)}}$}}
& \multicolumn{1}{r}{\multirow{2}{*}{$\Delta^{\mt{S(3)}}$ }}& 
\multicolumn{1}{r}{\multirow{2}{*}{$A_{ki}^{\mt{LR}}$}}
& \multicolumn{1}{r}{\multirow{2}{*}{$\Delta^{\mt{LR}}$ }} &
\multicolumn{1}{r}{\multirow{2}{*}{$A_{ki}^{\mt{MR}}$}} &
\multicolumn{1}{r}{\multirow{2}{*}{$\Delta^{\mt{MR}}$ }} \\
&&&&&&&&\\
\hline
T&2.067&	-0.057&	2.089&	-0.079&2.089& -0.079&2.17&0.16\\
Q&1.971&	0.039&	1.994&	0.016&1.993& 0.017&2.39&0.38\\
                \hline
	\end{tabular}

\vspace{10pt}

	\begin{tabular}{l d d d d d d d d}
		\hline
                \hline
\multicolumn{9}{c}{\multirow{2}{*}{{\bf Ba} $6s^2-6s6p$ : $A_{ki}^{\mt{exp}} = 1.19(4)$ 
\cite{klose2002critically,sansonetti2005handbook}}}\\
&&&&&&&&\\
		\hline
                \multicolumn{1}{c}{\multirow{2}{*}{}} &
 \multicolumn{1}{r}{\multirow{2}{*}{$A_{ki}^{\mt{S(2)}}$ }}
 & \multicolumn{1}{r}{\multirow{2}{*}{$\Delta^{\mt{S(2)}}$}} &
 \multicolumn{1}{r}{\multirow{2}{*}{$A_{ki}^{\mt{S(3)}}$}}
& \multicolumn{1}{r}{\multirow{2}{*}{$\Delta^{\mt{S(3)}}$ }}&
 \multicolumn{1}{r}{\multirow{2}{*}{$A_{ki}^{\mt{LR}}$}}
& \multicolumn{1}{r}{\multirow{2}{*}{$\Delta^{\mt{LR}}$ }} &
\multicolumn{1}{r}{\multirow{2}{*}{$A_{ki}^{\mt{MR}}$}} &
\multicolumn{1}{r}{\multirow{2}{*}{$\Delta^{\mt{MR}}$ }} \\
&&&&&&&&\\
\hline

T&1.285&	-0.095&	1.295&	-0.105&1.290& -0.100&1.65&0.46\\
Q&1.312	&       -0.122&	1.324&	-0.134& 1.323& -0.133&1.81&0.62\\
E46&1.305&	-0.115&	1.319&	-0.129&1.312& -0.122&1.87&0.68\\
                \hline
	\end{tabular}

\end{center}
\end{table}

\subsubsection{Quadrupole transition probabilities}
Electric quadrupole transitions are difficult to observe 
due to the very long lifetimes of the atomic \ce{D} states. 
For closed-shell atoms only the calcium and barium atomic 
\ce{^1D} states are directly connected with the ground  
\ce{^1S} states through the E2 transition. 
For the calcium atom two measurements of the quadrupole 
transition probabilities  were reported\cite{fukuda1982oscillator, beverini2003measurementof} 
with error bars that exclude one the other. 
Thus, accurate theoretical determination can discriminate 
between the two measurements. 
For barium the (old) experimental result\cite{klimovskii1981measurement}
 with relatively large error bars does not agree with any
 theoretical determination.\cite{mccavert1974oscillator, bauschlicher1984theoretical, migdalek1990multiconfiguration} 
Thus, the present results will shed some light on the accuracy of the measurements and calculations.

For Ca, we computed the 
 $4s^2 -\ 3s^14s^1$ quadrupole transition probability with the XCC3S(3) method in the def2-QZVPP basis set.\cite{weigend2005balanced}
The experimentally measured energy is $21849.63$ cm$^{-1}$.\cite{linstrom2013nist} 
As the energy in \frm{Dt-spring} and (\ref{Qt-short}) is present in third and fifth power, respectively,
 small error in the computed energy introduces a large 
error in the transition probability.
Therefore, we present the transition probabilities computed with both 
theoretical and experimental energy input.

\begin{table}[!ht]
\caption{Quadrupole transition probabilities for Ca. The XCC3
 and LRCC3 computations were performed in the cc-pV5Z basis set.\cite{peterson2002accurate, koput2002ab}}
\label{ca-res}
\begin{center}
	\begin{tabular}{l l l c r}
		\hline
                \hline
\multirow{2}{*}{$A$ $s^{-1}$} & \multirow{2}{*}{$S$} & \multirow{2}{*}{$E$} & \multirow{2}{*}{year} & \multirow{2}{*}{Ref.}\\
                &&&&\\
		\hline
	          $87$ &T&T&1980  & Ref.~\citenum{pasternack1980experimental}\\
	          $40\pm 8$ &E& E & 1982  & Ref.~\citenum{fukuda1982oscillator}\\
	          $81$ & T & T & 1981 & Ref.~\citenum{diffenderfer1981spin}\\
	          $39.6$&T&T&1985& Ref.~\citenum{bauschlicher1985radiative}\\
	          $60.2$ &T&T&1983 & Ref.~\citenum{beck1983electric}\\
	          $70.5$ &T&T&1991& Ref.~\citenum{vaeck1991mchf}\\
	          $54.4\pm 4$ &E& E &2003& Ref.~\citenum{beverini2003measurementof}\\
                  $49.42$ & T&T&2008& Ref.~\citenum{mitroy2008properties} \\
                  $66.44$ & T&T&2014& MRCI \\
                  $58.56$ & T&E&2014& MRCI \\
                  $56.08$ & T&T&2014& LRCC3 \\
                  $51.11$ & T&E&2014& LRCC3 \\
                  $56.05$ & T&T&2014 &XCC3S(3) \\
                  $51.08$ & T&E&2014 &XCC3S(3) \\
                \hline
                \hline
	\end{tabular}
\end{center}
\end{table}

\Frt{ca-res} shows the result for the
calcium E2 transition that have been published to date.
In the second and third columns, T stands for theoretically 
and E for experimentally obtained value for
the line strength and energy, respectively. The present 
theoretical results are well within the error bars of the
2003 measurement\cite{beverini2003measurementof} and 
outside the error bars of the older 1982 measurement.\cite{fukuda1982oscillator}
Note that the XCC3 and LRCC3 results are very close to each other 
despite quite different theoretical approaches that are on
the basis of these methods. Thus, we can conclude that
the present study supports the experimental result from 2003.\cite{beverini2003measurementof}

We also computed the quadrupole transition probabilities for the calcium 
atom with the MRCI method as this approach is based on a different model 
of the wave function. The results obtained with both the theoretical 
and experimental excitation energies are outside the error bars of the 
experiment from 2003. However, the value of the quadrupole transition 
probability calculated with the experimental excitation energy differs 
only by 1\% from the experimental result of \citet{beverini2003measurementof} 
which confirms once more that the experimental result from 2003 is more probable.

There are only a few theoretical values 
\cite{mccavert1974oscillator, bauschlicher1984theoretical, migdalek1990multiconfiguration}
 for the  $6s^2 - 6s5d$ transition in 
Ba, and only one experimental result.\cite{klimovskii1981measurement} 
The experimental 
transition energy is equal to  $11395.35$ cm$^{-1}$.
\cite{linstrom2013nist} 
We have employed the ECP46MDF pseudopotential and the corresponding $spdfg$ 
basis.\cite{lim2006relativistic, krych2011sympathetic}  \Frt{ba-res} compiles the  published
results for the $6s^2 - 6s5d$ Ba quadrupole transition. 
None of the earlier
theoretical results as well as the present XCC3 and LRCC3 results, 
are within the experimental error.
One should notice though that the experimental value error bars show a
 huge uncertainty.
The MRCI transition probabilities, 
both with the theoretical and experimental 
excitation energies, are also far from the experimental value. 
Note also that for the Ba atom the MRCI results are significantly different from both the LRCC3 and XCC3 results.

\begin{table}[!ht]
\caption{Quadrupole transition probabilities for barium.}
\label{ba-res}
\begin{center}
	\begin{tabular}{l l l c r}
		\hline
                \hline
\multirow{2}{*}{A $s^{-1}$} & \multirow{2}{*}{S }& \multirow{2}{*}{E }& 
\multirow{2}{*}{year} &\multirow{2}{*}{ Ref.}\\
                &&&&\\
		\hline
 	          $3.2$ &T&T&1974  & Ref.~\citenum{mccavert1974oscillator}\\
 	          $2.98$ &T&T&1984  & Ref.~\citenum{bauschlicher1984theoretical}\\
	          $3.381$ & T&T &1990  & Ref.~\citenum{migdalek1990multiconfiguration}\\
 	          $3.880$ &T&E&1990  & Ref.~\citenum{migdalek1990multiconfiguration}\\
                  $8\pm 3$ &E&E&1981  & Ref.~\citenum{klimovskii1981measurement}\\                
                  $2.47$ & T&T&2014& MRCI \\
                  $1.42$ & T&E&2014& MRCI \\
                  $3.49$ & T&T&2014 &LRCC3 \\
                  $2.85$ & T&E&2014 &LRCC3 \\
                  $3.52$ & T&T&2014 &XCC3S(3) \\
                  $2.87$ & T&E&2014 &XCC3S(3) \\
                \hline
                \hline
	\end{tabular}
\end{center}
\end{table}

\section{Conclusions}\label{concl}
We have presented an extension of the  coupled cluster 
 method
designed for the computation of the ground
state properties and transition probabilities. 
In order to test the performance of our method, 
we have computed  dipole moments for several
diatomic molecules.
The results were compared to the experimental data. 
A comprehensive analysis showed that the best compromise
between accuracy and computational 
cost is achieved for the XCC3S(3) variant, i.e.
for the third-order approximation to the auxiliary
operator.

We have reported the expressions 
for the transition density matrices computed from the
Hermitian formulation of the polarization propagator in  
 the XCC3 approximation.
In contrast to the LRCC3 method, the correct 
time-reversal symmetry of the line strength is 
guaranteed  by the algebraic construction of the polarization propagator 
in the XCC theory and its approximate variants.

The results of the transition probabilities 
computed with both the XCC3 and LRCC3 methods are of the same quality, 
though XCC is computationally less demanding.
The same conclusion holds for the XCC3 and LRCC3 dipole moments.

The computed dipole and quadrupole transition probabilities were
compared with the experimental data, and in most cases the results were
in a perfect agreement with the experiment.
Our results for the quadrupole transition probabilities in the
calcium atom with both the XCC3 and LRCC3 methods strongly favor the
new measurement of 2003.\cite{beverini2003measurementof} Our results
for the Ba atom are consistent with all the other theoretical data, suggesting
that the experimental determination should be reconsidered. 

The code for 
transition moments from the ground state will be incorporated 
in the KOŁOS: A general purpose ab initio program for 
the electronic structure calculation with Slater orbitals, 
Slater geminals, and Kołos-Wolniewicz functions.

\section*{Acknowledgment}
This work was supported by the Polish Ministry of Science 
and Higher Education within the grants No. NN204 215539 and  NN204 182840. 
RM thanks the
Foundation for Polish Science for support within the MISTRZ program.
\appendix*{}
\section{Biorthonormal, nonredundant basis for the triply excited manifold}
The general bra and ket vectors in the triply exited manifold are denoted as
$\langle ^{a_1a_2a_3}_{i_1i_2i_3} |$ and $ |^{a_1a_2a_3}_{i_1i_2i_3}\rangle$,
where the sequence of virtual-occupied electron pair indices is decreasing from
left to right.
In the case where all indices are different ($a_1>a_2>a_3$ and $i_1>i_2>i_3$)
the biorthonormal set  is defined  as :
\equsl{&v_1=|^{a_1a_2a_3}_{i_1i_3i_2}\rangle, \qquad v_2=|^{a_1a_2a_3}_{i_2i_1i_3}\rangle, \qquad
v_3=|^{a_1a_2a_3}_{i_2i_3i_1}\rangle,\\
&v_4=|^{a_1a_2a_3}_{i_3i_1i_2}\rangle,\qquad v_5=|^{a_1a_2a_3}_{i_3i_2i_1}\rangle,\\
&\widetilde{v}_1=
\frac{\langle^{a_1a_2a_3}_{i_1i_3i_2}|}{4} + \frac{\langle^{a_1a_2a_3}_{i_2i_1i_3}  |}{12} +
\frac{\langle^{a_1a_2a_3}_{i_2i_3i_1}  |}{6}+ \frac{\langle^{a_1a_2a_3}_{i_3i_1i_2}  |}{6}+
\frac{\langle^{a_1a_2a_3}_{i_3i_2i_1}  |}{12},\\
&\widetilde{v}_2=
\frac{\langle^{a_1a_2a_3}_{i_1i_3i_2}|}{12} + \frac{\langle^{a_1a_2a_3}_{i_2i_1i_3}  |}{4} +
\frac{\langle^{a_1a_2a_3}_{i_2i_3i_1}  |}{6} + \frac{\langle^{a_1a_2a_3}_{i_3i_1i_2}  |}{6}+
\frac{\langle^{a_1a_2a_3}_{i_3i_2i_1}  |}{12},\\
&\widetilde{v}_3=
 \frac{\langle^{a_1a_2a_3}_{i_1i_3i_2}|}{6} +
\frac{\langle^{a_1a_2a_3}_{i_2i_1i_3}  |}{6} + \frac{\langle^{a_1a_2a_3}_{i_2i_3i_1}  | }{3}+
\frac{\langle^{a_1a_2a_3}_{i_3i_1i_2}  |}{6}+ \frac{\langle^{a_1a_2a_3}_{i_3i_2i_1}  |}{6},\\
&\widetilde{v}_4=
\frac{\langle^{a_1a_2a_3}_{i_1i_3i_2}|}{6} + \frac{\langle^{a_1a_2a_3}_{i_2i_1i_3}  |}{6} +
\frac{\langle^{a_1a_2a_3}_{i_2i_3i_1}  |}{6} + \frac{\langle^{a_1a_2a_3}_{i_3i_1i_2}  |}{3}+
\frac{\langle^{a_1a_2a_3}_{i_3i_2i_1}  |}{6},\\
&\widetilde{v}_5=
\frac{\langle^{a_1a_2a_3}_{i_1i_3i_2}| }{12}+ \frac{\langle^{a_1a_2a_3}_{i_2i_1i_3}  |}{12}
+ \frac{\langle^{a_1a_2a_3}_{i_2i_3i_1}  | }{6}+ \frac{\langle^{a_1a_2a_3}_{i_3i_1i_2}  |}{6}+
\frac{\langle^{a_1a_2a_3}_{i_3i_2i_1}  |}{4}.
}{a3}

The vectors in \fr{a3} satisfy $\braket{\widetilde{v}_k}{v_l} = \delta_{kl}$.
Note that in this case there are only five linearly independent bra/ket vectors.
If some of the indices are equal, there are three cases to consider:
\begin{enumerate}
\item A single equality among the occupied indices (either $i_1 = i_3$ or $i_2 = i_3$)
\equ{\langle\widetilde{^{a_1a_2a_3}_{i_1i_2i_3}}| = \frac13\langle^{a_1a_2a_3}_{i_1i_2i_3}| + \frac16\langle^{a_1a_2a_3}_{i_2i_1i_3}|
}
\item A single equality among the virtual indices  (and an additional constraint on the occupied indices: $\neg (i_1 > i_2 > i_3)$)
\equ{\langle\widetilde{^{a_1a_2a_3}_{i_1i_2i_3}}| = \frac13\langle^{a_1a_2a_3}_{i_1i_2i_3}| + \frac16\langle^{a_1a_2a_3}_{i_3i_2i_1}|.
}
\item A single equality among the occupied indices and among the virtual ones 
(the equalities are indicated by repeating labels; additionally, the strict inequalities $a_1>a_2$ and $i_1>i_2$ hold)
\equs{&\langle\widetilde{^{a_1a_1a_2}_{i_1i_2i_1}}| = \frac12\langle^{a_1a_1a_2}_{i_1i_2i_1}|, \quad
\langle\widetilde{^{a_1a_1a_2}_{i_1i_2i_2}}| = \frac12\langle^{a_1a_1a_2}_{i_1i_2i_2}|, \quad\\
&\langle\widetilde{^{a_1a_2a_2}_{i_2i_1i_2}}| = \frac12\langle^{a_1a_2a_2}_{i_2i_1i_2}|, \quad
\langle\widetilde{^{a_1a_2a_2}_{i_1i_1i_2}}| = \frac12\langle^{a_1a_2a_2}_{i_1i_1i_2}|.
}
\end{enumerate}
All vectors that do not fit into the above defined templates are deemed 
linearly dependent and discarded from the basis. Note that this is one of possible
choices of the biorthonormal nonredundant basis.

\bibliography{ref}

\end{document}